\documentclass{article}
\usepackage[dvips]{graphicx}
\begin{document}
\title{Oscillatory phase transition and pulse propagation in noisy integrate-and-fire neurons}
\author{Hidetsugu Sakaguchi\\
Department of Applied Science for Electronics and Materials,\\
 Interdisciplinary of Engineering Sciences,\\
 Kyushu University, Kasuga Fukuoka 816-8580, Japan}
\maketitle
We study non-locally coupled noisy integrate-and-fire neurons with the Fokker-Planck equation. A propagating pulse state and a wavy state appear as a phase transition from an asynchronous state. We also find a solution in which traveling pulses are emitted periodically from a pacemaker region. 
\\
\\
PACS number: 05.45.Xt, 87.10.+e, 84.35.+i\\
\\
\\

Coherent oscillations are observed in neural systems such as the visual cortex and the hippocampus. The synchronization of the oscillators is considered to play important roles in neural information processing \cite{rf:1}. 
There are mainly two viewpoints in the research of the oscillatory activity in  neural systems.  In the first viewpoint, the activity of each neuron is expressed by the firing rate, and the coherent oscillation appears owing to the interaction of the excitatory and inhibitory neurons. 
Wilson-Cohen and Amari found first oscillatory behavior theoretically in interacting neurons \cite{rf:2,rf:3}. Recently, Robinson, Rennie and Rowe proposed a more elaborate model to explain various EEG rhythms and epileptic seizures \cite{rf:4}. If the spatial freedom is taken into consideration, the excitation wave can propagate. Wilson-Cowan performed numerical simulations of two layers of excitable neurons and inhibitory neurons \cite{rf:5}. 
In the second viewpoint, each neuron is regarded as an oscillator. Coherent oscillation appears as the global synchronization of the coupled oscillators.  
 The global synchronization in general coupled oscillators was first studied by Winfree \cite{rf:6}. Kuramoto proposed a globally coupled phase oscillator model as a solvable model for the global synchronization \cite{rf:7}.  The leaky-integrate-fire model is one of the simplest models for a single neuron and often used to study dynamical behaviors of neural networks.  Each neuron receives an input via synaptic connections from other neurons and it fires when the input goes over a threshold and sends out impulses to other neurons.  In that sense, the coupling is instantaneous, and then the model is called pulse-coupled oscillators.  Mirollo and Strogatz studied a globally coupled system of the integrate-and-fire neurons, and showed that perfect synchronization occurs in a finite time \cite{rf:8}. The synchronization of pulse coupled oscillators has been studied in deterministic systems by many researchers \cite{rf:9,rf:10,rf:11}. 
If each oscillator's behavior is stochastic, the model is generalized to a noisy phase oscillator model and a noisy integrate-and-fire model. 
In the stochastic system, the coherent oscillation appears as an analogue of the phase transition in the statistical mechanics. 
Globally coupled noisy phase oscillators were studied in \cite{rf:12,rf:13,rf:14,rf:15}, and globally coupled noisy integrate-and-fire model were studied in \cite{rf:16,rf:17,rf:18}. The globally coupled system is a useful model for the detailed analyses,  however, local or non-local interactions are more plausible, since neurons interact with other neurons via long axons or gap junctions. The non-locally coupled system of the deterministic integrate-and-fire neurons was also studied \cite{rf:19}. 
 In this paper, we study a non-locally coupled noisy integrate-and-fire model with the direct numerical simulation of the Fokker-Planck equation. 

The equation for a noisy integrate-and-fire neuron is written as 
\begin{equation}
\frac{dx}{dt}=1-bx+I_0+\xi(t),
\end{equation}
where $x$ is a variable corresponding to the membrane potential, $b$ is a positive parameter, $I_0$ denotes an external input, and $\xi(t)$ is the Gaussian white noise satisfying $\langle \xi(t)\xi(t^{\prime})\rangle=2D\delta(t-t^{\prime})$.  If $x$ reaches a threshold 1, $x$ jumps back to 0. If $b<1+I_0$, each neuron fires spontaneously. The Fokker-Planck equation for the Langevin equation (1) is 
\begin{equation}
\frac{\partial P}{\partial t}=-\frac{\partial}{\partial x}(1-bx+I_0)P(x)+D\frac{\partial^2P}{\partial x^2}+\delta(x)J_0(t),
\end{equation}
where $J_0(t)=-D(\partial P/\partial x)_{x=1}$ is the firing rate.  
The stationary distribution $P_0(x)$ for the Fokker-Planck equation (2) is written as \cite{rf:20}
\begin{eqnarray}
P_0(x)&=&P_0(0)e^{\{(ax-(1/2)bx^2\}/D},\;\; {\rm for}\; x<0, \nonumber\\
&=&P_0(0)e^{\{(ax-(1/2)bx^2\}/D}\left [1-\frac{\int_0^xe^{\{-az+(1/2)bz^2\}/D}dz}{\int_0^1e^{\{-az+(1/2)bz^2\}/D}dz}\right ],\;\;{\rm for}\;0<x<1,\nonumber\\
\end{eqnarray}
where $a=1+I_0$ and $P(0)$ is determined from the normalization condition $\int_{-\infty}^{1}P_0(x)dx=1$. The firing rate $J_0$ is determined as $J_0=DP_0(0)/\int_0^1e^{\{-az+(1/2)bz^2\}/D}dz$. We have performed direct numerical simulation of Eq.~(2) with the finite difference method with $\Delta x=0.0002$ and $\Delta t=2.5\times 10^{-5}$, and checked that the stationary probability distribution (3) is successfully obtained. 

\begin{figure}[htb]
\begin{center}
\includegraphics[width=13cm]{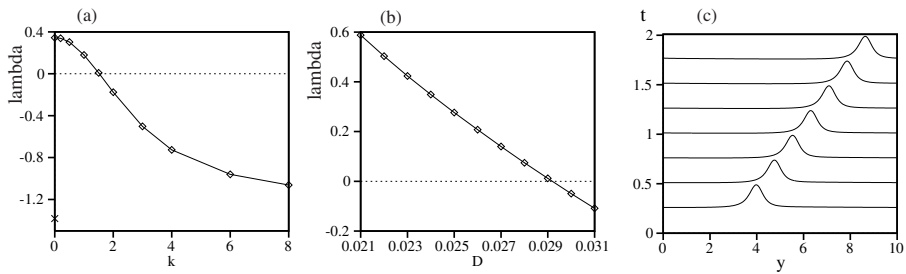}
\end{center}
\caption{(a) Linear growth rate $\lambda$ as a function of wavenumber $k$ of the uniform solution for $g_1(y)=1.2(1.5\exp(-4|y|)-0.1)$ at $D=0.025$. The cross mark on the vertical line indicates the linear growth rate for $k=0$. (b)  Linear growth rate $\lambda$ at $k=2\pi/L$ as a function of $D$. (c) Time evolution of the profiles of the firing rate $J_0(y)$.}
\label{fig:1}
\end{figure}
\begin{figure}[htb]
\begin{center}
\includegraphics[width=13cm]{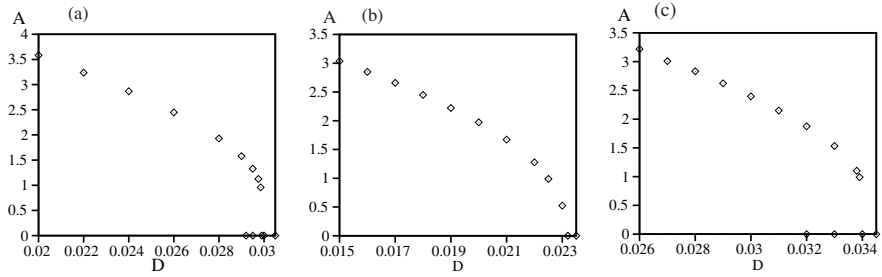}
\end{center}
\caption{Peak-peak amplitude $A$ of $I(y,t)$ as a function of $D$ for (a) $g_1(y)=1.2\{1.5\exp(-4|y|)-0.1\}$, (b) $g_1(y)=1.5\exp(-4|y|)-0.1$ and (c) $g_1(y)=1.3\{1.5\exp(-4|y|)-0.1\}$.}
\label{fig:2}
\end{figure}
\begin{figure}[htb]
\begin{center}
\includegraphics[width=13cm]{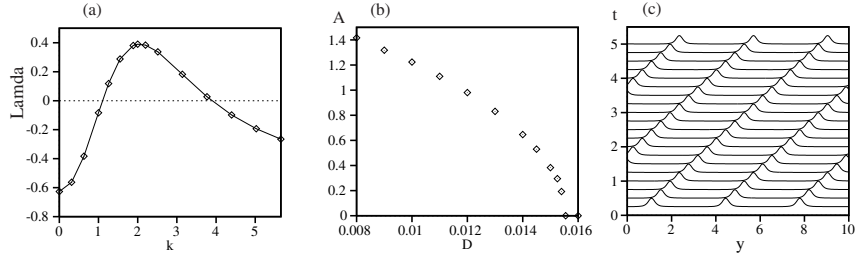}
\end{center}
\caption{(a) Linear growth rate $\lambda$ as a function of $k$ at  $D=0.01$ for $g_2(y)=1.2\{1.5\exp(-4|y|)-0.4\exp(-|y|)\}$. (b) Peak-peak amplitudes $A$ of $I(y,t)$ as a function of $D$. (c) Time evolution of the firing rate $J_0(y)$ for the pulse train state at $D=0.01$.}
\label{fig:3}
\end{figure}
\begin{figure}[htb]
\begin{center}
\includegraphics[width=6cm]{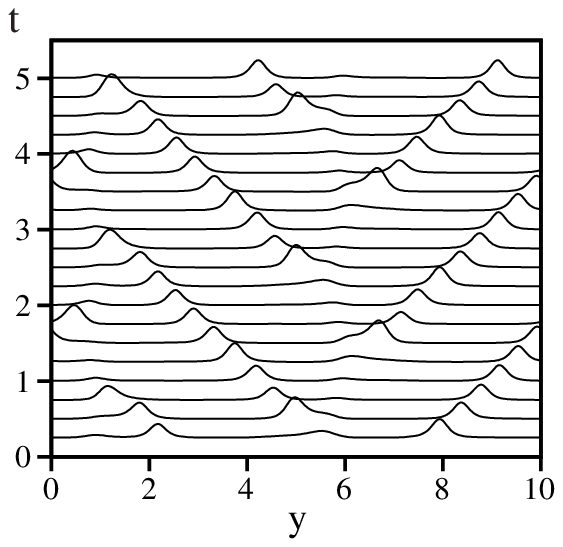}
\end{center}
\caption{Time evolution of the firing rate $J_0(y)$ for the wavy state with a pacemaker region at $D=0.01$.}
\label{fig:4}
\end{figure}
We assume a non-locally coupled system composed of the noisy integrate-and-fire neurons.  
Each neuron interacts with other neurons via synaptic connections. Time delay exists generally for the synaptic connections.  A model equation of the interacting noisy integrate-and-fire neurons is written as
\begin{equation}
\frac{dx_i}{dt}=1-bx_i+I_i+\xi_i(t),
\end{equation}
where $x_i$ denotes the dimensionless membrane potential for the $i$th neuron, $\xi_i(t)$ denotes the noise term which is assumed to be mutually independent, i.e., $\langle \xi_i(t)\xi_j(t^{\prime})\rangle=2D\delta_{i,j}\delta(t-t^{\prime})$,  and $I_{i}$ is the input to the $i$th neuron by the mutual interaction. The input $I_{i}$ to the $i$th neuron from the other neurons is given by 
\begin{equation}
I_i=\sum_{j}\sum_k g_{i,j}\frac{1}{\tau}e^{-(t-t_k^j)/\tau},
\end{equation}
where $t_k^j$ is the time of the $k$th firing for the $j$th neuron, $g_{i,j}$ denotes the interaction strength from the $j$th neuron to the $i$th neuron, and $\tau$ denotes a decay constant.  The sum is taken only for $t>t_k^j$.  The effect of the firing of the $j$th neuron to the $i$th neuron decays continuously with $\tau$. If $\tau\rightarrow 0$, the coupling becomes instantaneous.  Equation (5) is equivalent to 
\begin{equation}
\frac{dI_i}{dt}=-\{I_i-\sum_{j}\sum_kg_{i,j}\delta(t-t_k^j)\}/\tau.
\end{equation} 
If there are infinitely many neurons at each position $y$, we can define the number density of neurons with membrane potential $x$  clearly at each position. The number density is expressed as $n(x,y,t)$ at position $y$ and time $t$. The non-locally coupled system can be studied with a mean-field approach. In the mean-field approach, the number density is proportional to the probability distribution $P(x,y,t)$ for the probability variable $x$.  
The average value of $\delta(t-t_k^j)$ expresses the average firing rate at time $t$ at the position $y$. It is expressed as $J_0(y,t)=-D(\partial n/\partial x)_{x=1}$.  
The number density $n(x,y,t)$ therefore obeys the Fokker-Planck type equation: 
\begin{eqnarray}
\frac{\partial n(x,y)}{\partial t}&=&-\frac{\partial}{\partial x}(1-bx+I(y,t))n(x,y)+D\frac{\partial^2n}{\partial x^2}+\delta(x)J_0(y,t),\nonumber\\
\frac{dI(y,t)}{dt}&=&-\{I(y,t)-J(y,t)\}/\tau,\nonumber\\
J(y,t)&=&\int g(y,y^{\prime})J_0(y^{\prime},t)dy^{\prime},
\end{eqnarray}
where $g(y,y^{\prime})$ is the coupling  strength from the neuron located  at $y^{\prime}$ to the one at $y$, and $I(y),J_0(y)$ are respectively the input and the firing rate for the neuron at $y$. 

We have assumed that the time delay for the signal to transmit  between $y^{\prime}$ and $y$ can be neglected and $g(y,y^{\prime})$ depends only on the distance $|y-y^{\prime}|$, i.e., $g(y,y^{\prime})=g(|y-y^{\prime}|)$.  As two simple examples of the non-local coupling, we use $g_1(y,y^{\prime})=c\exp(-\kappa|y-y^{\prime}|)-d$ and $g_2(y,y^{\prime})=c\exp(-\kappa|y-y^{\prime}|)-d\exp(-\kappa^{\prime}|y-y^{\prime}|)$. These forms of the coupling imply that the interaction is excitable locally, but the interaction strength decreases with the distance $|y-y^{\prime}|$, and it becomes inhibitory when $|y-y^{\prime}|$ is large. This Mexican-hat type of coupling was used in several neural models \cite{rf:21}, especially to study the competitive dynamics in neural systems. Although two layer models of excitatory neuron layer and inhibitory neuron layer may be more realistic, we consider the above simpler one-layer model. The inhibitory interaction approaches a constant value $-d$ for the coupling $g_1$, and 0 for the coupling $g_2$. The system size is assumed to be $L=10$ as a simple example, and the periodic boundary conditions for the space variable $y$ are imposed. We choose the damping constants $\kappa$ and $\kappa^{\prime}$, as the exponential function decays to almost 0 for the distance $|y-y^{\prime}|\sim L$. Therefore, the dynamical behaviors do not depend on the system size $L$ qualitatively in the second model. But the dynamical behaviors depend on the system size $L$ in the first model, because the range of the inhibitory interaction is infinite in the model. 

There is a stationary and uniform solution $n(x,y,t)=n_0(x)$ and $I(y,t)=I_0$ in the non-locally coupled equation. The uniform solution satisfies 
\begin{eqnarray}
n_0(x)&=&n_0(0)e^{\{ax-(1/2)bx^2\}/D},\;\; {\rm for}\; x<0, \nonumber\\
&=&n_0(0)e^{\{ax-(1/2)bx^2\}/D}\left [1-\frac{\int_0^xe^{\{-az+(1/2)bz^2\}/D}dz}{\int_0^1e^{\{-az+(1/2)bz^2\}/D}dz}\right ],\;\;{\rm for}\;0<x<1,\nonumber\\
\end{eqnarray}
where the parameter $a$ is determined by the self-consistent condition
\begin{equation}
a=1-g_0D(\partial n_0(x)/\partial x)_{x=1},
\end{equation}
where $g_0=\int g(y,y^{\prime})dy^{\prime}$.

To study the linear stability of the stationary and uniform solution, we consider small deviations $\delta n(x,y,t)=n(x,y,t)-n_0(x)$ and $\delta I(y,t)=I-I_0$ from the uniform solution. The small deviations can be expressed with the Fourier series as $\delta n(x,y,t)=\sum \delta n_k(x,t)\exp(iky)$ and $\delta I(y,t)=\sum \delta I_k\exp(iky)$ under the periodic boundary conditions, where $k=2\pi m/L$. The perturbations $\delta n_k$ and $\delta I_k$  obey coupled linear equations 
\begin{eqnarray}
\frac{\partial \delta n_k(x,t)}{\partial t}&=&-\frac{\partial}{\partial x}\{(1-bx+I_0)\delta n_k(x,t)+\delta I_k(t)n_0(x)\}+D\frac{\partial^2\delta n_k}{\partial x^2}+\delta(x)\delta J_0(t),\nonumber\\
\frac{d\delta I_k(t)}{dt}&=&-\{\delta I_k(t)-g^{\prime}\delta J_{0k}(t)\}/\tau,
\end{eqnarray}
where $\delta J_{0k}(t)=-D(\partial n_k/\partial x)_{x=1}$ and 
$g^{\prime}=\int g(y,y^{\prime})e^{ik(y^{\prime}-y)}dy^{\prime}$. For $L$ is sufficiently large, $g^{\prime}=2c\kappa/(\kappa^2+k^2)-dL\delta_{k,0}$ for the coupling $g_1$ and $g^{\prime}=2c\kappa/(\kappa^2+k^2)-2d\kappa^{\prime}/(\kappa^{\prime 2}+k^2)$ for the coupling $g_2$. The stability of the stationary state is determined by the real part of the eigenvalues of the linear equation (10). But, the stationary solution $n_0(x)$ is a nontrivial function of $x$, and it is not so easy to obtain the eigenvalues. Here we have evaluated the real part of the largest eigenvalue of the linear equation for various $k$ by direct numerical simulations of Eq.~(10). 
The dynamical behavior in the long time evolution of Eq.~(10) is approximately expressed with the 
largest eigenvalue $\lambda$, that is, $\delta n_k$ and $\delta I_k\sim e^{\lambda t}$ for $t\gg 1$.   
We have numerically calculated the linear growth rate of the norm $\{\int (\delta n_k)^2dx+(\delta I_k)^2\}^{1/2}$ (which grows as $e^{({\rm Re} \lambda) t}$ for $t\gg 1$) every time-interval 0.001. 
Since the norm grows to infinity or decays to zero in the natural time evolution of the linear equation, we have renormalized the variables every time-interval 0.001, as the norm is 1 by the rescaling $c \delta n_k\rightarrow \delta n_k$ and $c \delta I_k\rightarrow \delta I_k$ with a constant $c$.  
We have regarded the average value of the linear growth rate of the norm as the largest eigenvalue for Eq.~(10).
Figure 1(a) displays the linear growth rate $\lambda$ as a function of $k$ for the coupling $g_1(y)=1.2\{1.5\exp(-4|y|)-0.1\}$ at $D=0.025$. The other parameters $b$ and $\tau$ are fixed to be $b=0.8,\,\tau=0.01$. There is discontinuity at $k=0$ for this coupling. The linear growth rate at $k=0$ takes a negative value denoted by the cross. The uniform state is stable for the uniform perturbation with $k=0$. The growth rate decreases with $k$, but it is positive for $k<2$. Figure 1(b) displays the linear growth rate as a function of $D$ for the coupling $g_1$ at $k=2\pi/L$. The uniform state is unstable for $D<0.0291$. The Hopf bifurcation occurs for a nonzero wavenumber.  Therefore, a wavy state is expected to appear for $D<0.0291$. We have performed direct numerical simulations for this coupling at $D=0.028$. Figure 1(c) displays a time evolution of the profiles of the firing rate $J_0(y,t)$. The profile of the firing rate has a pulse structure and it is propagating in the right direction. Since the pulse propagates one round $L$ with period $T=3.23$, the velocity of the traveling pulse is $L/T\sim 3.1$. A regular limit cycle oscillation with period $T$ is observed at each point. The directions depend on the initial conditions.  The traveling pulse state is an ordered state in the non-locally coupled system. The locally excitable interaction facilitates the local synchronization of the firing, but the global inhibition suppresses the complete synchronization. 
As a result of the frustration, a traveling pulse appears.  The pulse state is different form the traveling pulse observed in an excitable system, since the uniform state is unstable in our system and the pulse state is spontaneously generated from the stationary asynchronous state. 

The input $I(y,t)$ to the neuron at position $y$ exhibits regular limit cycle oscillation.  Figure 2(a) displays the peak-peak amplitude $A$, which is defined as the maximum value minus the minimum value of $I(y,t)$,  as a function of $D$. The oscillatory state disappears at $D=0.02985$ and the traveling pulse state changes into the stationary and uniform state. Inversely, as $D$ is decreased, the traveling pulse state appears spontaneously from the stationary state at $D=0.0291$, which is the critical value obtained from the linear stability analysis. That is, the phase transition is weakly subcritical for this coupling. We have changed the coupling function as $g_1(y)=\alpha(1.5\exp(-4|y|)-0.1)$ with a free parameter $\alpha$, and studied the phase transition at two other values of $\alpha=1$ and $\alpha=1.3$. The critical values $D_c$ by the linear stability analysis are $D_c=0.0232$ for $\alpha=1$ and $D_c=0.0320$ for $\alpha=1.3$. Figures 2(b) and (c) display the peak-peak amplitude $A$ of $I(y,t)$ as a function of $D$ by the direct numerical simulation of Eq.~(7). The bifurcation is supercritical for $\alpha=1$ and it is subcritical for $\alpha=1.3$. The parameter range $\Delta D=0.02$ of the hysteresis region is larger for $\alpha=1.3$  than the parameter range $\Delta D=0.0075$ for $\alpha=1.2$.  There is a transition from the supercritical bifurcation to the subcritical bifurcation at a critical value slightly smaller than $\alpha=1.2$. 

As a second example, we consider a non-locally coupled system with the coupling function $g_2(y)=1.8\exp(-4|y|)-0.48\exp(-|y|)$. Figure 3(a) displays the linear growth rate $\lambda$ for the stationary and uniform state as a function of $k$ at $D=0.01$. The linear growth rate is a continuous function of $k$ and takes a maximum at $k\sim 2$. The linear growth rate takes the largest value at wavenumber $k=6\pi/10$ (,i.e., wavelength $L/3$ ) in our finite size system of $L=10$. The linear growth rate at $k=6\pi/10$ takes positive values for $D<D_c=0.0155$.  We have performed direct numerical simulations for various $D$'s. A wavy state with the finite wavenumber $k=6\pi/10$ appears in this non-local system for $D<D_c$. Figure 3(b) displays the peak-peak amplitude of $I(y,t)$ as a function of $D$. A supercritical phase transition occurs at $D\sim 0.0155$, which is also consistent with the linear stability analysis. Near the critical value, the amplitude of the oscillation is small and the wavy state seems to be sinusoidal.
As $D$ is decreased, the oscillation amplitude increases and the sinusoidal waves change into pulse trains gradually. 
Figure 3(c) displays the time evolution of the profile of the firing rate $J_0(y,t)$ at $D=0.01$. This pattern was obtained by decreasing $D$ stepwise from the sinusoidal wave state near the critical point.  The pulse number is three and it 
is consistent with the result of the linear stability analysis.  
The velocity of propagating pulse is $v=L/(3T)=1.67$, where $T=2.0$ is the period of oscillation at a fixed position. 
Figure 4 displays a time evolution of a different type of wavy state. 
This state was obtained in a numerical simulation at the same parameter $D=0.01$ as Fig.~3(c), starting from the uniform initial condition with small random perturbations.  Pulses are created periodically near $x\sim 6$ and they are propagating alternatively in different directions. The inversely-propagating pulses collide at $x\sim 1$ and they disappear.  Namely, there are a pacemaker region (a source region) and a sink region of traveling pulses in this solution. 
This type of wavy state including a pacemaker region and the simple pulse-train state are bistable.  

To summarize, we have studied the non-locally noisy integrate-and-fire model with the Fokker-Planck equation.   We have found that a traveling pulse appears as a result of oscillatory phase transitions. We found also a pulse-train state by changing the form of the interaction. The wavy states appear as a phase transition from a asynchronous state when the noise strength is decreased. We have investigated a one-dimensional system for the sake of simplicity of numerical simulations, but 
we can generalize the model equation to a two-dimensional system easily.  Our non-locally coupled integrate-and-fire model might be too simple,  however, the wavy state is one of the typical dissipative structures far from equilibrium. Therefore, the spontaneously generated waves might be observed as some kind of brain waves also in real neural systems. 

\end{document}